\documentclass[preprint]{aastex}

\slugcomment{Submitted to \emph{Astrophysical Journal}}

\shorttitle{Precipitating Condensation Clouds}
\shortauthors{Ackerman and Marley}

\begin{document}

\title{Precipitating Condensation Clouds in Substellar Atmospheres}

\author{Andrew S. Ackerman\altaffilmark{1} and Mark S. Marley\altaffilmark{2,3}}
\affil{NASA Ames Research Center, Moffett Field, CA 94035}

\altaffiltext{1}{email: ack@sky.arc.nasa.gov}
\altaffiltext{2}{email: mmarley@mail.arc.nasa.gov}
\altaffiltext{3}{Department of Astronomy, New Mexico
State University, Las Cruces NM 88003}

\begin{abstract}

We present a method to calculate vertical profiles of particle
size distributions in condensation clouds of giant planets and
brown dwarfs.  The method assumes a balance between turbulent diffusion
and sedimentation in horizontally uniform cloud decks.  Calculations for
the Jovian ammonia cloud are compared with results from previous methods.  An
adjustable parameter describing the efficiency of sedimentation allows the
new model to span the range of predictions made by previous models. 
Calculations for the Jovian ammonia cloud are consistent with
observations.  Example calculations are provided for water,
silicate, and iron clouds on brown dwarfs and on a cool
extrasolar giant planet.  We find that precipitating cloud decks
naturally account for the characteristic trends seen in the spectra
of L- and T-type ultracool dwarfs.

\end{abstract}


\section{Introduction}

The visual appearance and spectrum of every solar system body with an atmosphere
depends strongly upon the character and distribution of atmospheric
condensates.  This is particularly true for the giant planets where optically
thick cloud decks dominate the appearance of the planets at most continuum
wavelengths in both the reflected solar and the thermal infrared.
Condensates also play a role in controlling the 
spectra of at least some brown dwarfs and 
most extrasolar giant planets. Indeed one suggested
classification scheme (Sudarsky, Burrows \& Pinto 2000) 
for extrasolar planets hinges on the specific atmospheric condensates present.  
Yet despite the importance of condensates there exists no simple model for 
predicting the parameters most relevant to radiative transfer: the
vertical profile of condensate mass and its distribution over
particle size.  

Chemical equilibrium models (e.g., Lewis 1969; Fegley \& Lodders 1994)
predict which species are expected to condense in an 
atmosphere, yet they provide no guidance as to the expected particle sizes.
Other models (e.g., Rossow 1978; Lunine et al. 1989;
Carlson, Rossow, \& Orton  1988)
predict some parameters, but lack a simple, self-consistent
recipe for exploring the possible phase space in which clouds might exist. 

A single example motivates the need for cloud models in substellar atmospheres. 
With increasingly later spectral type the warm L-dwarfs become progressively
redder in their $J-K$ color
(e.g., Kirkpatrick et al. 1999; Mart{\'i}n et al. 1999; Fan et al. 2000). 
Spectral fitting and models
(e.g., Leggett, Allard, \& Hauschildt 1998; Chabrier et al. 2000;
Marley 2000) demonstrate 
that this is due to the progressive appearance of more silicate
dust in the cooling brown dwarf atmospheres.  Yet the cooler T-type brown
dwarfs, like Gliese 229B, have blue colors in $J-K$ (e.g., Leggett
et al. 1999; Tsvetanov et al. 2000).  The spectra and colors
of these cool brown dwarfs can only be fit 
by atmosphere models that assume the silicate dust has settled below the 
visible atmosphere (e.g., 
Allard et al. 1996; Marley et al. 1996; Tsuji et al. 1996).
Models in 
which the dust does not settle (Chabrier et al. 2000) produce T-dwarf
colors that are at least 2 to 3 magnitudes redder than observed. 
Marley (2000)
has demonstrated that a simple model in which all clouds are
a single scale-height thick can explain this behavior, but the assumed
distribution was prescribed rather than being calculated from any model physics.
Correct modeling of the atmospheres of cooling brown dwarfs 
and the ultimate assignment of an effective temperature scale to the L-dwarf
spectral sequence (e.g., Kirkpatrick et al. 1999; Basri et al. 2000)
requires a characterization of clouds.  The ideal model 
would have a small number of free parameters, predict the vertical
distribution and particle sizes of the condensates, and yet be simple enough to
be included into model atmosphere 
codes that iteratively search for self-consistent atmospheric structures.
No such ideal model yet exists. 

We aim to fill this void by presenting a simple model describing 
precipitating
clouds in substellar atmospheres.  We limit our treatment to condensation
clouds, and hence do not consider photochemically driven hazes likely to
appear in illuminated stratospheres.  
We depart from previous work by explicitly treating the 
downward transport of raindrops with sizes greater than that
predicted from the convective velocity scale.  Including rainfall
produces clouds of thinner vertical extent, which can better reproduce 
observations of Jupiter's ammonia cloud.  The resulting model
is general enough to be applied to iron and silicate clouds
appearing in brown dwarf atmospheres (e.g., objects with effective
temperatures, 
$T_{\rm eff} \sim 1500\ \rm K$) as well as the atmospheres of cool extrasolar 
giant planets ($T_{\rm eff} \sim 400\,\rm K$) in which water 
clouds dominate the atmosphere.  The few free parameters in the model
can produce clouds with dramatically different characteristics; ultimately
observations will constrain these parameters and hopefully provide
information on the underlying atmospheric dynamics and cloud physics.

In this paper we first summarize previous cloud modeling efforts, then describe
the new model.  We use the ammonia cloud of Jupiter as a framework for
describing the model physics and evaluating the model performance.  Finally
we illustrate model applications by considering water, silicate, and iron
clouds in the atmospheres of brown dwarfs and a cool extrasolar giant planet.

\section{Previous Models} \label{previous}

A great range of models have been used to represent clouds in the terrestrial
atmosphere, which vary in the complexity by which atmospheric dynamics and
cloud microphysics are treated.  The most detailed models simulate
three-dimensional cloud-scale motions and resolve the size distributions
of cloud droplets (and the aerosols on which they form) and treat the
interactions between dynamics, microphysics, and radiative transfer.  
The computational demands of such complex models limit their domain sizes
to a few kilometers in each dimension.  Present global-scale (general 
circulation) models greatly simplify the representation of clouds
by parameterizing cloud-scale motions as well as cloud microphysical
processes, and such simplifications lead to profound uncertainties in
climate predictions from their simulations.  Both types of models, as well
as a range of intermediate models, can be considered appropriate for
modeling the terrestrial atmosphere by virtue of the wealth of
observational data available to constrain them; whether or not the
unknowns in such models are uniquely constrained by the data constitutes
a debate beyond the scope of this study.

The relative scarcity of observational data for clouds in extraterrestrial
atmospheres is far less constraining.  Leading uncertainties include the
characteristics of atmospheric dynamics and the populations
of nuclei upon which cloud droplets form.  Hence, we consider it appropriate
to model extraterrestrial clouds using much simpler treatments.

Perhaps the simplest approach to modeling clouds is through a Lagrangian
parcel model, in which the base of a cloud appears where the adiabatic
cooling of an air parcel in an updraft results in saturation 
(ignoring any supersaturation associated with barriers to cloud
droplet formation).  Further cooling condenses vapor in excess of
saturation onto cloud particles.  The particles grow through condensation 
and coalescence until their sedimentation
velocities exceed the updraft speed, and then fall out of the
parcel.  A number of problems arise in the formulation of updraft parcel
models, among them: ignoring parcels in downdrafts, treating the mixing
between parcels, treating the source of condensates into a parcel due
to sedimentation from above, and determining updraft speeds.

Another simple approach, which we employ here, is through a one-dimensional
Eulerian framework, in which turbulent diffusion mixes a condensable vapor
upwards, while maintaining a constant mixing ratio (equivalently, mole fraction)
below the cloud.  Temperature and hence the saturation mixing ratio in the air
column decrease with altitude, and the cloud base again appears where the
saturation mixing ratio matches the sub-cloud mixing ratio. Above the cloud
base, turbulent diffusion works toward maintaining a constant total mixing
ratio ($q_{\rm t} = q_{\rm v} + q_{\rm c}$), which is the sum of the vapor
($q_{\rm v}$ = moles of vapor per mole of atmosphere) and
condensate ($q_{\rm c}$ = moles of condensate per mole of atmosphere)
mixing ratios, while sedimentation reduces
$q_{\rm t}$ by transporting condensate downward.  Note that by ignoring
horizontal variability, any differences between (cloudy) updrafts and
(potentially cloud-free) downdrafts are neglected.

A number of models for tropospheric condensation clouds have appeared
in the planetary and astrophysical literature. Here we summarize a
selection of them that contribute to the present work. 

\subsection{Lewis (1969)}

Lewis (1969) represents a foundation in the study of tropospheric
clouds in the giant planets. In that work the term ``precipitation'' is
used in the narrow sense used by chemists,
in which condensates appear where the
local saturation vapor pressure is exceeded by the actual vapor
pressure, rather than in the broader sense used by meteorologists,
which additionally denotes sedimentation of the condensates (hereafter
we use the term in this broader sense).
Although there is no mention
of sedimentation by Lewis, the treatment does imply certain
assumptions. Starting below the cloud base and working upwards, at
each computational level the Lewis model assumes that all the
condensate remains at the level where it appears. Considered in the
framework of a parcel in an updraft, the Lewis model assumes that
all condensate rains out with a fallspeed matching the updraft velocity.
Were sedimentation slower, condensate would be
transported upward (as discussed by Weidenschilling \& Lewis 1973);
were sedimentation faster, condensate would be transported downward.
Hence, the Lewis (1969) assumption regarding sedimentation is an
unstated compromise between those two possibilities.  

We implement the Lewis model by starting below the
cloud base (where $q_{\rm c} = 0$ and $q_{\rm v} = q_{\rm below}$) and
condensing all vapor in excess of saturation at each successive level upward:
\begin{eqnarray}
\label{lewis_qc}
q_{\rm c}(z) & = & \max( 0, q_{\rm v}(z-\Delta z) - q_{\rm s}(z) ) \\
\label{lewis_qv}
q_{\rm v}(z) & = & \min( q_{\rm v}(z-\Delta z), q_{\rm s}(z) ) 
\end{eqnarray}
where $z$ is altitude and $q_{\rm s}$ is the vapor mole fraction 
corresponding to the saturation vapor pressure at that altitude.
The first and second cases on the right
hand side correspond to cloud-free and cloudy conditions, respectively.
Note that under all conditions $q_{\rm t}(z) = q_{\rm v}(z-\Delta z)$ in
the Lewis model,
reflecting the assumption that only vapor is transported upwards.

Beyond this simple model, Lewis (1969) considered the partitioning of
chemical species in some detail, and also calculated pseudo-adiabatic
lapse rates.  Here we simply assume that each condensate results from
the saturation of a single condensable, and fix the lapse rate as input
from observations or an external model.

For an example, we calculate an ammonia cloud profile from the
Lewis model (Figure 1) using the Jovian temperature profile from
Voyager (Lindal et al. 1981), 
the relation for vapor pressure given in Appendix
\ref{pvap}, and a sub-cloud
mole fraction of 3~$\times$~$10^{-5}$ 
(a wide range of abundances below the expected base of the Jovian ammonia
cloud have been reported; we adopt the value at 0.6 bars retrieved by
Kunde et al. 1982 for the Northern Equatorial Belt, which also
agrees with the best-fit values of Carlson, Lacis, \& Rossow 1993
and Brooke et al. 1998).
The cloud base appears at 0.42 bars, where the temperature is 129 K. Although
absent in the figures of Lewis (1969) (likely due to reduced vertical
resolution), in our interpretation
of that model the vapor is not entirely depleted in the lowest reaches
of the cloud (where $q_{\rm c} < q_{\rm t}$), hence $q_{\rm c}$ increases 
above the cloud base.  Such an increase is found in terrestrial clouds of
moderate vertical extent, 
where $q_{\rm c} < q_{\rm t}$, and hence $q_{\rm c}$
increases with altitude throughout their depths. However,
at greater altitudes in this deep ammonia cloud, the vapor is so effectively
depleted by condensation at the low temperatures that
$q_{\rm v} \ll q_{\rm t}$, leading
to a cold degeneracy:
$q_{\rm c}(z) \approx q_{\rm t}(z) = q_{\rm s}(z-\Delta z)$, in which
decreasing temperatures result in $q_{\rm c}$ diminishing with altitude.
Note that the
condensate abundance drops off rapidly above $\sim$0.13 bars due to increasing
temperatures. Hence, the temperature minimum quite
reasonably produces a cold-trap in the Lewis model.

Condensate particle sizes, 
the other ingredient needed for predicting cloud opacity, 
are not considered by Lewis (1969) or Weidenschilling \& Lewis (1973).

\subsection{Carlson et al. (1988)}

In their theoretical characterization of cloud microphysics of the
giant planets, Carlson et al. (1988) employ the formalism of Rossow (1978)
to calculate time constants for droplet condensation within
cloudy updrafts (assuming a supersaturation of 10$^{-3}$), 
droplet coalescence (assuming the mean collision rate is described by
particles with a mass ratio of 2), and sedimentation through an
atmospheric scale height. From these time constants, estimates are made
of the predominant size of cloud particles at cloud base for a number
of condensates. For the Jovian ammonia cloud, Carlson et al. estimate
a mass-weighted droplet radius of $\sim 10 - 30 \: \mu$m.

Carlson et al. (1988) make no attempt to calculate vertical profiles of
condensate mass.  For profiles of vapors that condense into multiple
forms (such as ammonia, which can also condense onto a cloud of
NH$_4$SH below the ammonia cloud), saturation is assumed above the cloud
base.

A shortcoming to the approach of Carlson et al. (1988) 
is that their microphysical time constants strongly depend on a
number of uncertain factors, chief among them completely unknown
supersaturations, which govern droplet growth rates due to condensation. 
Supersaturations in a cloudy updraft are determined by balance between
the source due to adiabatic cooling, and the sink due to condensation.
Uncertainties in updraft speeds and the populations of
condensation nuclei (and hence cloud droplets) both contribute
to the uncertainty in supersaturations realized in 
extraterrestrial clouds.  
Furthermore, the time constants Carlson et al. (1988) use for gravitational
coalescence assume that the collection efficiency is unity, and
those for sedimentation effectively assume a fixed width of the size
distribution.  Rather than attempting to constrain the many degrees of
freedom using such a detailed approach, we choose instead to reduce the
number of assumptions by simplifying the description of cloud microphysics.

\subsection{Lunine et al. (1989)}

Lunine et al. (1989) consider a range of possible iron and
silicate clouds in brown dwarfs; the possibilities differ in the nature
of the balance between sedimentation and turbulent mixing.
The framework is based on a theoretical investigation into
iron clouds deep in the Jovian atmosphere by Prinn \& Olaguer (1981),
which in turn draws on an analysis of sulfuric acid clouds on Venus
(Prinn 1974). These models represent a fleshing out of the discussion
of vertical transport of condensates by Weidenschilling \& Lewis (1973). 

Two fundamental cloud types are treated by Lunine et al. (1989). The first
is ``dust-like'' (using the terminology of Prinn \& Olaguer 1981) in
which cloud particles grow and efficiently sediment out, resulting in
relatively thin clouds limited by the local vapor pressure, as in the model
of Lewis (1969). These dust-like clouds are assumed to prevail in the
radiative region (stratosphere), where the temperature profile is stable
and convection is suppressed.  

The second fundamental type in the Lunine et al. (1989) study is a
tropospheric cloud, in which downward transport by sedimentation is
opposed by upward transport due to turbulent mixing.
For this cloud type, two variations are considered by
Lunine et al. (1989). For the first variation, described as
``frozen-in,'' cloud particles are so small that sedimentation is
overwhelmed by upward transport due to turbulent mixing. In this
case, the atmosphere is well-mixed with respect to condensate, and
hence $q_{\rm c}$ is independent of altitude above the cloud base.
For the second variation, which is intermediate to the dust-like and
frozen-in cases, particles grow large enough in ``convective'' clouds 
to develop appreciable sedimentation velocities, and their
downward sedimentation is balanced by their turbulent transport upward.
For their calculations of specific brown dwarf models, Lunine et al. (1989)
consider only the two endmembers of their cloud spectrum,
corresponding to dust-like and frozen-in clouds.

Their intermediate case serves as a starting point for our model
of condensate mass profiles.
Our interpretation of the convective cloud model of Lunine et al. (1989) 
as applied to the Jovian ammonia cloud is shown in Figure 1. Note
that we have refined that model slightly, allowing the atmospheric
properties to vary with height above the cloud base, and relaxing their
assumption that $q_{\rm c}$ = $q_{\rm t}$. The condensate mass is
seen to be significantly enhanced above the cloud base for that model:
at the tropopause (where there is no cold-trap in this case)
$q_{\rm c}$ is enhanced a thousand-fold over that computed by the
Lewis (1969) model.  Thus the treatment in which Lunine et al. (1989) 
assume particle sedimentation to balance turbulent transport results
in a cloud not so different from their frozen-in case (as depicted
by the curve in
Figure 1 labeled $f_{\rm rain} = 0$).  Evidently the sedimentation
in this convective cloud model is far less effective than that assumed by
Lewis (1969).  As described below, for our calculations
of condensate mass profiles we extend the Lunine et al. (1989) approach
by applying a scale factor to the particle sedimentation.

For radiative calculations, Lunine et al. (1989) assume all
particles in the frozen-in and dust-like clouds are 1 and 10 $\mu$m in
radius, respectively.

\subsection{Marley et al. (1999)}

The Marley et al. (1999) model of water and silicate clouds in
extrasolar giant planets represents a variation on the Lewis (1969)
model. As in the Lewis model, the calculation of vapor pressure
(or equivalently, $q_{\rm v}$) assumes that any supersaturation is quenched
locally by condensation (Equation~\ref{lewis_qv}).
However, the calculation of the
condensate mole fraction ($q_{\rm c}$) represents a departure: instead of
calculating it from the vapor pressure in the underlying layer from
Equation~\ref{lewis_qc}, Marley et al. scale it to the local saturation
vapor pressure with the following assumption:
\begin{equation}
q_{\rm c}(z) = \cases{0 & if $q_{\rm v}(z-\Delta z) < q_{\rm s}(z)$ \cr
                f_{\rm s} q_{\rm s}(z) & otherwise \cr }
\end{equation}
The parameter $f_{\rm s}$ corresponds to the potential supersaturation prior to
condensation. Marley et al. (1999) treat $f_{\rm s}$ as an adjustable parameter,
ranging from a baseline value of 0.01 to an extreme value of 1. The
baseline model as applied to the Jovian ammonia cloud is shown in
Figure 1. The condensate mass is seen to be diminished by a factor of
$\sim$100 relative to the Lewis (1969) model.

Increasing $f_{\rm s}$ to 1 results in a hundredfold enhancement of $q_{\rm c}$
throughout the cloud compared to the baseline case, as shown in Figure
2\emph{a}. The principle difference between that extreme and 
the Lewis (1969) condensate model is that for the former there is no
regime near the cloud base akin to shallow terrestrial clouds,
in which $q_{\rm c}$ increases with altitude.  This difference is attributable
to a discontinuity of $q_{\rm t}$ in the treatment of Marley et al.:
below the cloud base
$q_{\rm t}(z) = q_{\rm v}(z-\Delta z)$ as in the Lewis (1969) model,
but above the cloud base
$q_{\rm t}(z) = (1+f{\rm s}) \: q_{\rm s}(z)$.

For their calculations of cloud particle sizes, which are decoupled
from their calculation of condensate mass, Marley et al. (1999) apply the
formalism of Rossow (1978) to two atmospheric endmembers: first, a
quiescent atmosphere, in which the mean particle size is determined
from the condition that the sedimentation rate matches the faster of
coagulation and condensation (at an assumed supersaturation of 0.01);
and second, a turbulent atmosphere in which mixing is balanced by
sedimentation. 

The first endmember is subject to a similar catalog of unconstrained
assumptions as required by the treatment of Carlson et al. (1988),
the most notable among them being the great uncertainty in the
supersaturation driving droplet condensation.
Also, the model physics underlying this quiescent atmosphere seems to be
self-contradictory, on the one hand explicitly assuming that there is too little
convection to regulate the maximum size of the droplets, yet on the other
hand implicitly assuming that there is enough convection to supply
the vapor necessary to drive condensational growth.

However, the second case of Marley et al. (1999) requires significantly fewer
assumptions, and is also appropriate to tropospheric condensation clouds.
This second case serves as a starting point for the calculation of the
cloud particle sizes in our model.

\section{The Present Model}

We model all condensation clouds as horizontally homogeneous 
(globally averaged) structures 
whose vertical extent is governed by a balance between the upward
turbulent mixing of condensate and vapor
($q_{\rm t} = q_{\rm c} + q_{\rm v}$) and the downward transport
of condensate due to sedimentation:
\begin{eqnarray}
\label{K_pde}
-K \frac{ \partial q_{\rm t}}{\partial z} - f_{\rm rain} w_* q_{\rm c} = 0
\end{eqnarray}
where $K$ is the vertical eddy diffusion coefficient and $f_{\rm rain}$ is 
a new parameter that we have introduced,
defined as the ratio of the mass-weighted droplet sedimentation velocity
to $w_*$, the convective velocity scale. 
We solve Equation~\ref{K_pde} for each condensate independently, and hence
ignore any microphysical interactions between clouds.
Equation~\ref{K_pde} is an extension of Lunine et al.'s convective cloud
model, relaxing their implicit assumptions $f_{\rm rain}$ = 1 and 
$q_{\rm c} = q_{\rm t}$. 

The product $f_{\rm rain} w_*$ represents an average sedimentation
velocity for the condensate, which offsets turbulent mixing and thereby
leads to $q_{\rm t}$ decreasing with altitude.
The extreme case with no sedimentation to offset
turbulent mixing ($f_{\rm rain}$ = 0)
is equivalent the frozen-in endmember of
Lunine et al. (1989) and the ``dusty'' models
of the Lyon group (e.g., Chabrier et al. 2000).
In this case the solution to Equation~\ref{K_pde} is a well-mixed
atmosphere ($q_{\rm t}$ independent of altitude), which is seen in Figure 1
to loft even more condensate than the convective cloud of
Lunine et al. (1989).  

We adopt $f_{\rm rain}$ as an adjustable input parameter,
which together with $q_{\rm c}$ constrains the droplet size distributions. 
First we describe our calculation of $q_{\rm c}$, then the size distributions.

\subsection{Condensate Mass Profiles}

The eddy diffusion coefficient ($K$) for $q_{\rm t}$ is assumed to be the
same as that for heat as derived for free convection
(Gierasch and Conrath, 1985):
\begin{eqnarray}
\label{K_equals}
K = \frac{H}{3} \left( \frac{L}{H} \right)^{4/3}
    \left( \frac{R F}{\mu \rho_{\rm a} c_{\rm p}} \right)^{1/3}
\end{eqnarray}
where the atmospheric scale height is given by $H = R T / \mu g$ (for
Jupiter we use $g$~=~25~m~s$^{-2}$), $L$ is the turbulent mixing length,
$R$ the universal gas constant,
$\mu$ the atmospheric molecular weight (2.2 g mol$^{-1}$ assumed here),
$\rho_{\rm a}$ the atmospheric density, and $c_{\rm p}$ the specific heat
of the atmosphere at constant pressure (ideal gas assumed).
Here we assume all the interior heat to be transported through the
convective heat flux:
$F = \sigma T_{\rm eff}^4$,
where $\sigma$ is the Stefan-Boltzmann constant and the effective temperature
for Jupiter is $T_{\rm eff}$ = 124 K.  In the general case, a profile of $F$ is
specified by an external model, which partitions the transport of interior
heat between radiative and convective fluxes.  The convective heat flux
can be reduced further by other heat fluxes, such as para hydrogen
conversion or latent heat release, as discussed by Gierasch and Conrath (1985).
Beyond any uncertainty in the convective heat flux, the constant coefficient
scaling the eddy diffusion coefficient is only loosely constrained by
observations.  For our baseline model we use a coefficient of 1/3 based on
previous modeling studies of the Jovian atmosphere (D. M. Hunten 1996,
private communication), and consider the sensitivity of our model results
to its value in a subsequent section.

For freely convecting atmospheres, the mixing length is typically assumed
to be the pressure scale height.
However, in stable atmospheric regions, the mixing length will be 
diminished.  We account for this reduction by scaling the mixing length
to the local stability:
\begin{eqnarray}
\label{L_equals}
L = H \max( \Lambda, \Gamma/\Gamma_{\rm adiab})
\end{eqnarray}
where $\Gamma$ and $\Gamma_{\rm adiab}$ are the local and dry adiabatic
lapse rates, respectively, and $\Lambda$ is the minimum scaling applied to
$L$ (we assume a value of 0.1).
In the general case, convective heat fluxes are diminished in
radiative regions; hence, we also assume an eddy diffusion
coefficient no less than a prescribed minimum value ($K_{\rm min} = 
10^5$ cm$^2$ s$^{-1}$ in our baseline model), which
represents residual turbulence due to breaking buoyancy
waves (Lindzen 1981) and such.
A list of prescribed/adjustable parameters is provided in Table 1.

The remaining parameter in Equation~\ref{K_equals} is the convective
velocity scale from mixing-length theory: $w_* = K / L$.
Our baseline values for the turbulent mixing parameters just below the
Jovian ammonia cloud are $H = L = 20 \rm \, km$,
$K = 2 \times 10^8 \rm \, cm^2 s^{-1}$, and
$w_* = 1.1 \rm \, m \, s^{-1}$.

To compute the vertical distributions of condensate and vapor, we proceed
upwards from the sub-cloud conditions, requiring all excess vapor
to condense and solving Equation~\ref{K_pde}
at each level.  If we heuristically assume that $q_{\rm c} / q_{\rm t}$
and $L$ are constant in a cloud, the solution is an exponential decline
of total mixing ratio with height above cloud base (where we define $z = 0$):
\begin{equation}
\label{heuristic_soln}
q_{\rm t}(z) = 
q_{\rm below} \exp\left( - f_{\rm rain}
\frac{q_{\rm c}}{q_{\rm t}}
\frac{z}{L} \right)
\end{equation}
Note that by using the sub-cloud mixing ratio as a lower boundary
condition, any moistening due to rain evaporating 
below the cloud base is ignored. 

Comparing our adaption of the Lunine et al. (1989) profile with our calculation
using $f_{\rm rain} = 1$ isolates the effect of reducing the mixing length
due to atmospheric stability.
The cumulative effect of the progressive reduction in mixing length due
to the stability of the Voyager temperature profile above the cloud base
is seen to result in a cold-trap in the lower stratosphere.
Tripling $f_{\rm rain}$ further reduces the cloud density and
lowers the cold-trap to the tropopause; 
increasing it to 10 results in a cloud with less condensate
than the Lewis (1969) model. 

We assume that $f_{\rm rain}$ is independent of altitude. Yet specifying an
appropriate value of $f_{\rm rain}$ at the cloud base, let alone any vertical
dependence, poses a significant challenge. For guidance, first we turn to
\emph{in situ} measurements and detailed simulations of terrestrial water
clouds, and then consider constraints provided by values retrieved through
remote sensing of Jovian ammonia clouds.

For terrestrial stratocumulus clouds capping well-mixed planetary
boundary layers, we find that $f_{\rm rain} < 1$ in the cloud deck
and increases with distance below cloud top. 
An assortment of \emph{in situ} measurements indicates that
$f_{\rm rain}$ increases with decreasing droplet concentrations ($N$),
as fewer and hence larger droplets more efficiently 
produce drizzle and thereby decrease cloud water.  For example, 
$f_{\rm rain} \sim$ 0.2 for a case study over the
North Sea, where $N$ = 100 cm$^{-3}$ (Nicholls, 1984), while 
in California stratocumulus it increased from 0.3 to 0.5
as $N$ decreased from 40 cm$^{-3}$ in clouds contaminated by
ship exhaust, to 10 cm$^{-3}$ in clean ambient air
(from Figure 3\emph{b} of Ackerman et al. 2000b). 
We note that in clean
marine stratocumulus, reduced rain is observed (e.g., Taylor
and Ackerman 2000) and predicted (e.g., Ackerman, Toon, and Hobbs 1993;
Stevens et al. 1998) to result in deeper cloud layers, which is
consistent with our simple model.  However, any changes in cloud cover
associated with precipitation changes are not represented in our model.

We have also calculated profiles of $f_{\rm rain}$ for deeper convection,
using large-eddy simulations of trade cumulus clouds (Ackerman et al. 2000a),
which are twice as deep as stratocumulus clouds ($\sim$1000 m
compared to $\sim$500 m). As in the stratocumulus clouds, we find that
$f_{\rm rain} < 1$ in the stratiform anvils at cloud top.  However,
$f_{\rm rain}$ is significantly enhanced throughout the bulk of the
trade cumulus clouds, where it ranges from $\sim$ 2 to 6. 
Vertical winds are more symmetrically distributed in stratocumulus,
and hence the convective velocity scale is representative of the mean
updraft velocity. The circulation
is more skewed in trade cumulus, with narrow updrafts opposing the broad
subsidence: for the Ackerman et al. (2000a) trade-cumulus simulations the
mean vertical velocity in
cloudy updrafts is three times the convective velocity scale.
Such a skewness is consistent with an enhancement of the rain factor
in deep convection.

We recommend a more systematic analysis of $f_{\rm rain}$
in terrestrial clouds.  Perhaps more pertinent to the much deeper
clouds expected in gas giants and brown dwarfs, we also recommend
consideration of much deeper convection than considered here.  
For now we treat $f_{\rm rain}$ as an adjustable parameter, leaning toward
values $>$ 1, as we expect the deep convection in substellar
atmospheres (on the order of an atmospheric scale height and deeper) to
more closely resemble cumuliform than stratiform convection in the
terrestrial atmosphere.  

For observations of Jovian ammonia clouds, we first turn to the retrievals
obtained from the Voyager IRIS (InfraRed Interferometer Spectrometer)
instrument by Carlson, Lacis, \& Rossow (1994), who considered latitudinal
variations among zones and belts in the Jovian tropics. Although they did not
provide profiles of condensed ammonia, we can compare our results
to their ratios of condensate to atmospheric scale height
($H_{\rm p}/H_{\rm g}$ in their notation).  
Carlson et al. (1994) retrieve ratios of 0.35 and 0.40 ($\pm$ 0.10)
for ammonia clouds in the Equatorial and Northern Tropical Zones,
respectively, which are seen in Figure 3 to span a range of $f_{\rm rain}$ 
values between $\sim$ 1 and 3.  Retrievals of ammonia cloud properties for
the Jovian tropics from ISO (Infrared Space Observatory) measurements 
by Brooke et al. (1998) indicate a scale height ratio of 0.3, which is
consistent with our model results for $f_{\rm rain} \sim 2$.

Were the ratio $q_{\rm c} / q_{\rm t}$ fixed in our model, as 
assumed heuristically for Equation~\ref{heuristic_soln}, the 
condensate height should vary as $f_{\rm rain}^{-1}$.  However,
as seen in Figure 3, the model dependence is not nearly that steep,
and for $f_{\rm rain} > 3$ the dependence nearly vanishes.  The dependence
is moderated by a negative feedback in which $q_{\rm c} / q_{\rm t}$ decreases
with increasing $f_{\rm rain}$.  Our assumption of zero
supersaturation within a cloud is equivalent to
$q_{\rm c} / q_{\rm t} = S - 1$, where $S$ is the potential supersaturation
before condensation eliminates it.
Increased $f_{\rm rain}$ offsets more of the turbulent mixing of
vapor and condensate, thereby reducing the potential supersaturation
and decreasing the ratio $q_{\rm c} / q_{\rm t}$.  The negative feedback
is thereby due to the reduction of $q_{\rm c} / q_{\rm t}$, which 
diminishes the dependence of the condensate scale height on $f_{\rm rain}$
(Equation~\ref{heuristic_soln}).

The assumption that all vapor in excess of saturation condenses can
be relaxed in our model by replacing Equation~\ref{lewis_qc} with 
\begin{equation}
q_{\rm c}(z) = \max( 0, q_{\rm v}(z-\Delta z) -
              ( S_{\rm cloud} + 1) \: q_{\rm s}(z) )
\end{equation}
where $S_{\rm cloud}$ is the supersaturation that persists after accounting
for condensation.  Allowing $S_{\rm cloud} > 0$ represents 
conditions in which there is a significant barrier to the formation of
cloud droplets, and/or condensation is too slow to effectively offset
the supersaturation driven by cooling in updrafts. In shallow
terrestrial water clouds, such as stratocumulus, neither of these
conditions holds, as there are typically abundant condensation nuclei
upon which droplets form at low supersaturations, and the
concentration and diffusivity of water vapor are sufficient to allow
condensation to balance the modest dynamic forcing at low supersaturations
($\sim$10$^{-3}$).
However, in
cirrus clouds that form directly from the vapor phase (as opposed to
the freezing of water droplets from deep convection), there are
typically few effective ice nuclei available, and hence the barrier to
nucleation can result in supersaturation building to $\sim$0.5 before
ice crystals form, even in moderate updrafts. High supersaturations
($\sim$0.3) can be maintained after nucleation in cirrus clouds because
the concentration and diffusivity of water vapor are greatly reduced at
the cold temperatures of the upper troposphere (Jensen et al. 2000).

As extreme cases, in Figure 2\emph{b} the condensate profile for
$S_{\rm cloud}$ = 0 is compared to that for $S_{\rm cloud}$ = 1.
The barrier to condensation
results in a lifting of the cloud base and enhanced lofting of vapor, but
the altitude of the cold-trap is unchanged (recall that
the temperature profile is fixed here).
The greater lofting enhances condensate mass above $\sim$0.4
bar, which tends to increase the column of condensate. However,
the tendency is more than offset by the decrease in atmospheric mass
density at the elevated cloud base, and hence the condensate column
decreases from
52 to 39 g m$^{-2}$ in response to increasing $S_{\rm cloud}$ from 0 to 1.
This 25$\%$ reduction
of condensate column contrasts markedly with the effect of
increasing $f_{\rm s}$ from
0.01 to 1 in the Marley et al. (1999) model, in which the condensate
column increases a hundredfold (Figure 2\emph{a}). These
opposite responses (to comparable
changes in maximum relative humidities: from 100 to 200$\%$ for the
present model,
and 100.01 to 200$\%$ for the Marley et al. model) arise from the distinct
definitions of the supersaturation factor: $S_{\rm cloud}$ corresponds to the
supersaturation \emph{after} condensation is treated in the present model,
whereas $f_{\rm s}$ corresponds
to the potential supersaturation \emph{before} condensation is treated in the
Marley et al. model.  Marley et al.'s $f_{\rm s}$ is more comparable to
$f_{\rm rain}^{-1}$ than to $S_{\rm cloud}$ in the present model.

\subsection{Droplet Size Distributions}

Size spectra of cloud particles in terrestrial condensation clouds
are commonly observed as bimodal number distributions, with
a condensational mode (of radius $\sim$10 $\mu$m in stratocumulus)
resulting from
condensational growth at modest supersaturations, and a precipitation mode
at larger radii resulting from coalescence due to dispersion in sedimentation
velocities.  The bimodal structure
is evident in droplet size distributions measured in stratocumulus clouds
off the coast of California, as shown in Figure 4\emph{a}.
At the cloud base, the mean size of the condensation mode depends on
factors such as the updraft velocity and the composition and size
distribution of the condensation nuclei upon which droplets form. Above
the cloud base, the mean size of the condensation mode increases with altitude
in an updraft. 

The production of precipitation in terrestrial clouds is generally
observed to increase with the mean size and spectral width of the
condensation mode.  
The mean size of droplets in the precipitation mode\footnote{
The maximum size
of precipitation particles is set by their breakup due to hydrodynamic
instability, which for terrestrial raindrops occurs at a radius of
$\sim$3 mm.}
is found to increase with distance from cloud top in marine stratocumulus
(e.g., Nicholls 1984) (the opposite tendency of the condensation
mode), typically explained in microphysical terms as due to ``fortunate''
collector drops sweeping up smaller droplets and growing as they fall.
Such a profile is consistent, as it must be, with the observed decrease
in $f_{\rm rain}$ with height above the cloud base.  
We note in passing that measured profiles of $f_{\rm rain}$ are
also consistent with our assumption that the size
of precipitation particles decreases as convective velocity decreases
above the cloud base (approaching the temperature inversion that caps the
boundary layer),
though we assume a uniform value of $f_{\rm rain}$ in our model calculations.

We make no attempt to model the complexity of cloud 
processes here, as such detailed computations are prohibitively
demanding, and the parameter space of unknowns is overwhelming 
for the range of condensates expected in substellar atmospheres
(including such basic issues as whether condensates are solid or
liquid).  Instead, we simply prescribe a single, broad lognormal
size distribution of condensate particles at each level
(Figure 4), thereby halving the number of parameters required for
a bimodal distribution.  The lognormal size distribution is given by
\begin{equation}
\frac{dn}{dr} = \frac{N}{r \sqrt{2\pi} \ln \sigma_{\rm g}}
         \exp\left( -\frac{ \ln^2(r/r_{\rm g})}{2\ln^2 \sigma_{\rm g}} \right)
\end{equation}
where $n$ is the number concentration of particles smaller than radius $r$,
and the three parameters to be constrained appear on the right side:
$N$ is the total number concentration of particles, $r_{\rm g}$
the geometric mean radius, and $\sigma_{\rm g}$ the geometric
standard deviation.

Marley et al. (1999) also prescribe a lognormal
size distribution of condensate particles, in which
$\sigma_{\rm g}$ is fixed at 1.5 and
$r_{\rm g}$ is determined from the particle sedimentation
velocity corresponding to the convective velocity scale ($w_*$).
We choose to use $\sigma_{\rm g}$ as an adjustable parameter,
and determine $N$ and $r_{\rm g}$ through $q_{\rm c}$ and $f_{\rm rain}$.
The rain factor is rigorously defined through
\begin{eqnarray}
\label{frain_equals}
f_{\rm rain} = \frac{\int_0^\infty v_{\rm f} \: (dm/dr) \: dr}
                {\epsilon \rho_{\rm a} w_* q_{\rm c}}
\end{eqnarray}
where $v_{\rm f}$ is the particle sedimentation velocity (described in 
Appendix \ref{vfall}), $m$ is particle mass, and $\epsilon$ is the
ratio of condensate to atmospheric molecular weights. 
To close the system analytically we fit a power-law dependence for
particle fallspeed about its value at 
$v_{\rm f}( r_{\rm w} ) = w_*$ through
\begin{equation}
v_{\rm f} = w_* ( r / r_{\rm w} )^\alpha
\end{equation}
where the exponent $\alpha$ is calculated from a fit to the fallspeeds
between $r_{\rm w}/\sigma$ and $r_{\rm w}$ when $f_{\rm rain} > 1$, and
between $r_{\rm w}$ and $r_{\rm w} \, \sigma$ otherwise 
($\sigma$ is constrained to be $\geq 1.1$ for the fit). 
The power-law approximation allows Equation~\ref{frain_equals}
to be expressed as
\begin{eqnarray}
\label{frain_analytic}
f_{\rm rain} = \frac{\int_0^\infty r^{3+\alpha} \: (dn/dr) \: dr}
                {r_{\rm w}^\alpha \int_0^\infty r^3 \: (dn/dr) \: dr }
\end{eqnarray}
Integration of the lognormal distribution then leads to
\begin{eqnarray}
\label{rg_equals}
r_{\rm g} = r_{\rm w} f_{\rm rain}^{1/\alpha}
\exp\left (-\frac{\alpha+6}{2} \ln^2\sigma_{\rm g} \right)
\end{eqnarray}
\begin{eqnarray}
\label{N_equals}
N = \frac{3 \epsilon \rho_{\rm a} q_{\rm c}}{4 \pi \rho_{\rm p} r_{\rm g}^3}
\exp \left( -\frac{9}{2} \ln^2 \sigma_{\rm g} \right)
\end{eqnarray}
where $\rho_{\rm p}$ is the density of a condensed particle (see Appendix
\ref{vfall}).
The parameter $f_{\rm rain}$
can be interpreted in terms of microphysics by identifying the radius
of mass-weighted sedimentation flux in Equation~\ref{frain_analytic}:
\begin{eqnarray}
r_{\rm sed} = \left( 
          \frac{\int_0^\infty r^{3+\alpha} \: (dn/dr) \: dr}
                {\int_0^\infty r^3 \: (dn/dr) \: dr }
          \right)^{1/\alpha}
\end{eqnarray}
which leads to $f_{\rm rain} = ( r_{\rm sed} / r_{\rm w})^\alpha$.
In Figure 4\emph{b} the size distribution 
is weighted by precipitation, where it is
seen that $r_{\rm sed} < r_{\rm w}$, consistent with $f_{\rm rain} < 1$.
In contrast, for the large-eddy simulations of trade cumulus mentioned
above, and as implied by the retrievals from the Jovian ammonia clouds,
the droplets grows sufficiently large to satisfy
$r_{\rm sed} > r_{\rm w}$. 

It is seen from Equations \ref{frain_analytic}$-$\ref{N_equals} that
droplet sizes are decoupled from condensate mass (though both 
depend on $f_{\rm rain}$);
the condensate mass simply scales the distribution through $N$.
Ignoring any vertical dependence of atmospheric stability 
(Equation~\ref{L_equals}), 
vertical variations in droplet sizes are due to the height dependence of
convective velocity, leading to $r_{\rm w} \propto \rho_{\rm a}^{1/3\alpha}$,
which yields a mild vertical dependence of approximately
$r_{\rm w} \propto \rho_{\rm a}^{1/4}$ for our baseline Jovian ammonia cloud
(for which $\alpha = 1.3$, corresponding to a moderately turbulent sedimentation
regime in which fallspeeds are reduced from those in viscous flow).

Of greater interest than the mild vertical dependence are the
sensitivities of droplet sizes on $f_{\rm rain}$ and $\sigma_{\rm g}$ at
the base of our Jovian ammonia cloud, where $r_{\rm w} = 35~\mu$m.
For a given value of $\sigma_{\rm g}$, 
more efficient rain implies larger droplets:
$r_{\rm g} \propto f_{\rm rain}^{1/\alpha}$.
For our baseline droplet distribution (with $\sigma_{\rm g} = 2$),
a value of $f_{\rm rain} = 3$ leads to $r_{\rm g} = 14~\mu$m 
(compare to $\sim$ 10 $\mu$m for shallow terrestrial clouds),
while $f_{\rm rain} = 5$ yields $r_{\rm g} = 20~\mu$m.  

For efficient precipitation ($f_{\rm rain} > 1$), reducing the width of
the size distribution requires an increase in $r_{\rm g}$, as the narrower
distribution is centered on a larger radius.  For example, a
monodisperse distribution ($\sigma_{\rm g} = 1$) with $f_{\rm rain} = 3$
results in $r_{\rm g} = 77~\mu$m.
Such narrow distributions of large particles could result from
condensational growth at high supersaturations,
reminiscent of methane ``rain without clouds'' suggested by
Toon et al. (1988) for Titan's atmosphere.

In the complete model, we calculate spectrally-resolved profiles of
condensate opacity by integrating the scattering and absorption
coefficients (from Mie calculations) over the particle size distributions.
Here we simply present opacities for geometric scatterers, which for
a model layer of thickness $\Delta z$ is given by
\begin{equation}
\label{tau_equals}
\Delta\tau = \frac{3}{2} \frac{ \epsilon \rho_{\rm a} q_{\rm c} }
                  { \rho_{\rm p} r_{\rm eff} } \Delta z
\end{equation}
where the effective (area-weighted) droplet radius is evaluated from the
lognormal size distribution:
\begin{equation}
\label{reff_equals}
r_{\rm eff} = r_{\rm w} f_{\rm rain}^{1/\alpha}
\exp \left(-\frac{\alpha+1}{2} \ln^2\sigma_{\rm g} \right)
\end{equation}
The effective radius can be greater or less than $r_{\rm w}$, depending
on the combination of $f_{\rm rain}$, $\alpha$, and $\sigma_{\rm g}$.
At the base of our baseline Jovian ammonia cloud
$r_{\rm eff} = 46~\mu$m, which is 11 $\mu$m greater than $r_{\rm w}$.

The computation of cloud optical depth depends on vertical grid
resolution due to the exponential temperature dependence of saturation
vapor pressures.  To reduce such resolution dependence we progressively
subdivide each model layer until its optical depth converges to
1\% precision.  For the Jovian ammonia cloud our calculations converge
at a minimum sub-layer thickness of $\sim$ 30 m.

Equations~\ref{tau_equals} and \ref{reff_equals} show that increased
precipitation reduces opacity not only by decreasing $q_{\rm c}$ but
also by increasing $r_{\rm eff}$.
For heuristic purposes the column optical depth can be estimated from
the cloud base properties by ignoring any height dependence of the
mixing length and atmospheric scale height and assuming
$q_{\rm c} = q_{\rm t}$ within the cloud,
in which case
\begin{equation}
\tau = \frac{3}{2} \frac{ \epsilon p q_{\rm below}}
            { g r_{\rm eff}(1+f_{\rm rain})}
\end{equation}
where $p$ and $r_{\rm eff}$ are atmospheric pressure and droplet effective
radius at the cloud base. Note that $\tau$ is more than linearly dependent
on $q_{\rm below}$, due to its dependence on the cloud base pressure.
For the following comparisons with observations of the Jovian ammonia cloud,
we sum optical depths from Equation~\ref{tau_equals} over the model layers
and compaute a cloud average $r_{\rm eff}$ from the cloud optical depth
and vertical column of condensate.

\section{Comparisons with Observations}

West et al. (1986) attempt to reconcile among a vast array
of observations of the Jovian ammonia cloud, and conclude that its
optical depth at visible wavelengths is between $\sim$ 2 and 10,
comprised of a population of small particles ($r \sim 1\:\mu$m)
reaching the tropopause in both belts and zones,
underlain in zones by a population of larger particles
($r \sim 3 - 100\:\mu$m) concentrated near the cloud base.  

More recently, Banfield et al. (1998), who make no attempt to retrieve
particle sizes, assume a particle effective radius of
0.2 $\mu$m and retrieve optical
depths from Galileo imaging data at a wavelength of
0.756 $\mu$m that cluster in the range $\sim1 - 4.5$ (their Figure 9). 
For comparison with the optical depths of ours and West et al.'s
(at a mid-visible
wavelength of 0.55 $\mu$m), we scale Banfield et al.'s $\tau$ by the ratio
of extinction efficiencies
at 0.55 to 0.756 $\mu$m for 0.2 $\mu$m particles (using their refractive
index of 1.4), resulting in a mid-visible $\tau$ range of $\sim2 - 10$.  Hence,
the West et al. (1986) and Banfield et al. (1998) results
are effectively identical, and are hereafter lumped together
as ``West et al.''

Recall from Figure 3 that the condensate scale height retrievals of
Carlson et al. (1994) are consistent with our baseline model for
$f_{\rm rain}$ between $\sim$ 1 and 3.  This entire
$f_{\rm rain}$ range for our baseline model (in which $\sigma_{\rm g} = 2$)
is seen in Figure 5{\it a} to overlap with the data reported by
West et al.  (1986). The overlap corresponds to $r_{\rm eff}$ ranging
from $\sim$ 18 to 45 $\mu$m.
A narrower size distribution results in larger droplet effective radii
and therefore smaller optical depths; for the monodisperse case 
the data mutually overlap at $r_{\rm eff} \sim$ 34 to
60 $\mu$m.

Carlson et al. (1994) use Voyager IRIS spectra
to retrieve cloud optical depths\footnote{
Carlson et al. (1994) report cloud optical depths for an assumed extinction
efficiency of 1, whereas by treating cloud particles as geometric scatterers,
we assume an extinction efficiency of 2.  Hence, we multiply their optical
depths by 2 for comparison with ours.}
and droplet sizes for the Northern Equatorial Belt `hot-spots'
(note that the infrared data are insensitive to the submicron particles
reported by West et al. 1986).
Their best-fit particle distribution 
is a mixture of small ($r_{\rm eff} = 3\:\mu$m,
$\tau$ = 0.16) and larger ($r_{\rm eff} = 100\:\mu$m, $\tau$ = 0.38)
particles, resulting in a combined $r_{\rm eff}$ = 72 $\mu$m
and $\tau$ = 0.54.
However, hot spots are anomalous features of reduced cloudiness associated
with pronounced dynamical forcings (e.g., Showman \& Ingersoll 1998),
and our one-dimensional model is
intended to represent horizontally averaged conditions, which would
seem more comparable to the Equatorial and Northern Tropical Zones, where 
Carlson et al. retrieve $\tau$ of $\sim 1.2$ and 2, respectively.
Carlson et al. do not present separate retrievals of droplet sizes for
the zones, but do state that their optical depths are
dominated by larger particles, which we interpret
as an observed range of $70\:\mu{\rm m} \leq r_{\rm eff} \leq 100\:\mu{\rm m}$.

Given that our model results overlap with the entire range of the 
West et al. (1986) optical depths, which in turn do not overlap with
those of Carlson et al. (1994), it should not be surprising
that there is no mutual overlap between our baseline model results and the
retrievals of $\tau$ and $r_{\rm eff}$ by
Carlson et al. (1994).  
However, our model results with $f_{\rm rain} = 3$ just overlap with the 
Carlson et al. (1994) retrievals if a monodisperse size distribution is
assumed, which suggests that the Carlson et al. data are dominated by
particles in the precipitation mode.
Alternatively, multiplying our eddy diffusion
coefficients by a factor of 3 also results in larger particles and therefore
reduced optical depths, which leads to overlap with the Carlson et al. (1994)
retrievals for $f_{\rm rain} \sim 2$ (Figure 5{\it b}).

The above comparisons show that reasonable choices of model parameters
produce agreement with observations of tropical Jovian ammonia clouds.
The few unknown model parameters in our simple model are not uniquely
constrained, befitting
the incompleteness and uncertainty in the observations and the ambiguity in
comparing a model for globally averaged clouds with measurements
that resolve large-scale horizontal variability in the clouds.  To
compare our model with observations averaged over a wider area
we next consider measurements of the Jovian tropics obtained from
Earth orbit.

Brooke et al. (1998) use a 3-$\mu$m ISO spectrum to retrieve 
microphysical properties of the Jovian ammonia cloud and find 
best fits for two possibilities: first, a monomodal distribution of
10-$\mu$m ammonia particles with a visible optical depth of 1.1; and
second, a bimodal distribution of 1 and 10-$\mu$m ammonia particles
with an optical depth of 1.3 (equivalent to an effective radius of 7 $\mu$m).
Both fits include an additional optical depth of 0.1 from grey 
particles.  In their analysis, a fit for 10-$\mu$m particles
indicates a superior fit in comparison to
those for 1 and 30 $\mu$m particles, which we intepret as
allowing a size range of 5 to 20 $\mu$m.  Recalling
from the discusson of Figure 3 that the Brooke et al. (1998) condensate
scale height (a single value with no uncertainty) is consistent
with $f_{\rm rain} = 2$.  Figure 5{\it b} indicates that mutual
consistency with the effective radii retrieved by Brooke et al. (1998) 
requires increasing the width of our size distrubutions and/or reducing
our eddy diffusion coefficients.  Simultaneously matching
the Brooke et al. (1998) optical depths requires a substantial reduction
of the sub-cloud ammonia abundance.  However, the baseline ammonia
abundance we use is reported  
by Brooke et al. (1998) (as a single value with no uncertainty)
to best fit their 3-$\mu$m spectrum.
Hence, our calculations are evidently inconsistent with that baseline
ammonia abundance and the combination of small particles and small
optical depths retrieved by Brooke et al. (1998).  

We have already noted the shortcoming that our model excludes the
possibility of horizontal variability.  For the case of modeling
emitted radiative flux, this simplification will of course result in
an underestimate at some wavelengths, since any flux leaking out through
the clearings between patchy clouds is not treated.  Horizontal
variability is also ignored in the retrievals of
Brooke et al. (1998),
which results in their underestimating cloud optical depth due to a
plane-parallel albedo bias (e.g., Calahan et al. 1994):
the area-weighted albedo of a cloud deck calculated from a 
single column with an optical depth $\tau$ is always greater 
than the albedo averaged over a variety of columns with the same 
area-weighted average $\tau$.  
Inverted for the Brooke et al. (1994) retrievals of optical depth 
from reflected spectra, this bias indicates that
the optical depth from an area-weighted average
radiance underestimates the area-weighted optical depth. 
Hence, the actual area-weighted optical depth is greater than
reported by Brooke et al. (1994) and closer to the optical
depths calculated by our model.  We are unable to quantify the
magnitude of this error without reproducing their retrievals, which
is beyond the scope of this study.

\section{Applications to Substellar Atmospheres}

\subsection{Vertical Cloud Structure}

We use our baseline model (Table 1) to calculate profiles of
condensed water,
silicate (as enstatite, MgSiO$_3$), and iron in theoretical atmospheres 
of brown dwarfs and a giant planet (see Figure 6 for gravities and effective
temperatures).
The temperature profiles are calculated for cloud-free conditions
(Marley 2000). 
The L-dwarf-like atmosphere (Figure 6\emph{a}) is too
warm for water to condense. 
Between 1 and 10 bars, within the convective region,
the silicate and iron clouds are seen to overlap, suggesting the 
possibility of microphysical interactions between them, which are
ignored by our model.  
The silicate particles in this case are about twice as large as the iron
particles because the assumed density of a silicate particle is
about half that of an iron particle (Appendix~\ref{vfall}).

Although the temperature profiles in Figure 6 were not 
calculated self-consistently to include the effects of clouds,
it is clear that
for objects near $T_{\rm eff} \sim 1500$~K, clouds are
an important opacity source.  The silicate and iron clouds in the
L-dwarf-like model (Figure 6\emph{a}) appear in
the visible atmosphere and therefore play an important role
in controlling opacity and the temperature structure of the
atmosphere.  Nevertheless these clouds are confined
to a relatively thin cloud deck, which does not reach the upper
regions of the atmosphere as do condensates in the well-mixed
profiles also shown.  The cloud particles are also fairly
large ($r_{\rm eff} \sim 40 - 80\,\rm \mu m$) and will have
a substantially different spectral opacity than smaller particles.
The Lyon group (Chabrier et al. 2000) employs an ``astrophysical
dust'' size distribution of sub-micron particles to model
dust opacity in such atmospheres.  The cloud model presented
here, with larger particles confined to a discrete cloud deck,
represents a substantial departure from the previous work.  In
a future publication we will discuss the spectral and color
properties of atmospheres with these new cloud models.

Figure 6\emph{b} presents the cloud model applied to a T-dwarf-like
atmosphere with $T_{\rm eff}=900\,\rm K$, which is again too warm
for water to condense. 
(No iron cloud is shown in Figures 6\emph{b} and \emph{c} because 
the cloud base is below the bottom of the model domain.)
Although the silicate cloud and the omitted iron cloud may be
important to the atmospheric temperature structure, they
no longer represent significant opacity sources to an observer.

The changing role of cloud opacity with effective temperature
is more clearly shown in Figure 7, which illustrates the brightness
temperature spectra of several 
radiative-equilibrium models for brown dwarf atmospheres as well as 
the atmospheric temperature range over which most of the cloud 
opacity is found.  In a model with $T_{\rm eff} = 1800\,\rm K$
(Figure 7{\it a}) the 
silicate cloud deck forms in the model stratosphere and is relatively 
thin.  Comparison of the solid and dotted curves, which respectively
include and exclude silicate and iron cloud opacities,
shows little difference between the two cases. 
Since the cloud optical depth is only a few tenths, flux is
efficiently transported from levels deeper than the base of the cloud. 
The iron cloud (not shown) adds a few more tenths of optical depth. 
Hence, the clear and cloudy models are very similar.  Such a model 
would be appropriate for an early type L-dwarf.  

Figure 7{\it b}
shows the results for a cooler atmosphere, with $T_{\rm eff} = 
1400\,\rm K$, appropriate for a late L-dwarf (Kirkpatrick et al. 
1999; Stephens et al. 2001).  Here the cloud is much more optically 
thick, and flux originates no deeper than the middle of the cloud 
layer.  In the clear atmosphere, flux originates from 
deeper, hotter levels.  As a result the band depths of the cloudy 
model are shallower, a result shown for dusty M-dwarfs by Jones and 
Tsuji (1997).  Since flux is conserved for $T_{\rm eff}$ fixed,
the cloudy model emits more 
flux beyond about $2\,\rm\mu m$ than the clear model. 
Regions of strong molecular absorption, including the
depths of the water bands shortwards of 2 $\mu$m as well
as most of the 2 to 5 $\mu$m region evidence higher
brightness temperatures in the cloudy case since the
atmosphere above the cloud must warm to produce the
same total emitted flux as in the cloud free
calculation.  Regions of stronger molecular opacities 
are sensitive to these warmer temperatures higher in
the atmosphere.

In Figure 7{\it c}, for which $T_{\rm eff} = 900\,\rm K$ the silicate
cloud forms well below the region in which most flux originates and again
the clear and cloudy models are similar.  However in the regions in
which the molecular 
opacity is lowest, near 1.1 and 1.3 $\mu m$, flux originates from
deeper regions in the clear atmosphere than for the cloudy case. 
As a result the peak-to-trough variation in emitted flux is again somewhat 
smaller for the cloudy model.  We note that clear atmosphere models 
for T-dwarfs like Gl229 B and GD165B typically over predict the 
water band depths (e.g., Marley et al. 1996; Allard et al. 1996;
Tsuji et al. 1996; Saumon et al. 2000; Geballe et al. 2001),
and suggest that the
attenuation of flux by the top of the silicate cloud deck may be
responsible for this effect.

Notably, these model atmospheres illustrate the origin of the curious
change in infrared colors of the L- and T-dwarfs 
(e.g., Kirkpatrick et al. 1999; Mart{\'i}n et al. 1999; Fan et al. 2000). 
The cloud-free cases shown 
in the figure monotonically vary in $J-K$ from 1.39 to -0.17 from 
warmest to coolest.  In contrast the cloudy models initially become 
redder with falling $T_{\rm eff}$ (moving from 1.6 to 1.7) before 
they move to the blue ($J-K = 0.38$ for the case in Figure {\it c})
when the silicate and iron
clouds begin to disappear below optical depth unity in the gas.
Thus our precipitating condensation cloud model qualitatively reproduces
the color variation of the L- and T-dwarfs, consistent with  
previous arguements based on interpretation of spectra (Allard et al. 1996; 
Marley et al. 1996; Tsuji et al. 1996).
In contrast, the pure chemical equilibrium model (Chabrier et al. 2000)
predicts the
presence of substantial dust opacity well to the top of the atmosphere,
which is clearly excluded by the data.
A more complete treatment of color changes
will be given in a future study.

For the cooler atmosphere representative of a cool extrasolar giant planet 
(Figure 6\emph{c}), water condenses in the radiative region, in essence
a stratospheric cloud. 
The ice particles are seen to be larger than the silicate particles,
chiefly due to the lesser densities assumed for the individual
ice particles.  The reduced gravity in the atmosphere of the
less massive extrasolar giant planet requires larger silicate particles to 
match the mean sedimentation velocity than does the more massive T-dwarf.
Note again that the well-mixed assumption produces a profoundly different
vertical structure, in which the silicate cloud is so deep that it
significantly overlaps the water cloud.

As did Marley et al. (1999), Sudarsky et al. (2000) have computed
water cloud profiles in order to estimate extrasolar giant
planet albedos.  Sudarsky et al. (2000) essentially assume
$f_{\rm rain}=0$ and limit the cloud to be no more than 1
scale height thick.  Such a model would be similar to
the well-mixed water cloud in Figure 6\emph{c} with a flat cloud top
at $\sim 4 \times 10^{-3}$ bars.

The emergence of water clouds in substellar atmospheres
with $T_{\rm eff}$ below about 500 K will reshape the
vertical temperature profile and emergent spectra of these objects.
Preliminary models computed with this cloud profile suggest that
such cool objects will again move to the red in $J-K$ after the
blueward excursion caused by the sinking of the silicate cloud
below the visible atmosphere and the emergence of $\rm CH_4$
as a dominant opacity source in $K$ band.  Hence the near-IR colors
of very cool objects computed from cloud-free atmosphere models
(e.g., Burrows et al. 1997) are likely to differ substantially
from actual objects.

\subsection{Non-uniform Clouds}

Variable brightness in $I$ band has been detected for some L-dwarfs by
Bailer-Jones \& Mundt (2001), who attribute the variability to evolution
of dust clouds.  They find some evidence that variability may be more
common in later-type L-dwarfs.  Although we do not model
horizontally variable clouds, these observations are consistent with
the model presented here.  As clouds form in progressively cooler
objects they become more optically thick and form deeper within the 
convective region of the atmosphere.  Thus global scale tropospheric 
weather patterns, as seen on Jupiter and predicted for 
brown dwarfs (Schubert \& Zhang 2000),
can more easily produce photometric variability since the turbulent
motions are greater, making local clearings more likely, and
enhancing the potential contrast between clear and cloudy air.
Indeed the great red spot of Jupiter produces a
photometric signal in both reflected sunlight and emitted thermal
radiation (Gelino \& Marley 2000).

Horizontally varying silicate clouds, even if not of the appropriate
scale to produce a varying photometric signal, may play an important
role in the transition from the dusty L-dwarfs to the relatively
cloud-free T-dwarfs.  
The change in $J-K$ color from the latest red L-dwarfs ($J-K\sim2$) to 
the blue T-dwarfs ($J-K\sim0$) is quite abrupt.  Four L8 dwarfs with 
known or estimated absolute magnitudes are only 1 magnitude brighter 
in $J$ band (Reid et al. 2001) than Gl229B.  Reid et al. argue that 
this implies the L8 dwarfs are only about 250 K warmer than Gl229B. 
Even with the silicate cloud deck forming at progressively deeper 
levels with falling $T_{\rm eff}$, it may be difficult to account 
for such rapid color variation.  In fact the rapid transition may 
be a signature of horizontally varying clouds.
Once tropspheric convective patterns begin to produce substantial
horizontal variability, the flux from the more cloud-free regions
will begin to dominate the total emitted flux, even if large fractions
of the object are still cloudy.  For example, Jupiter's 5-$\mu$m flux is
dominated by the relatively cloud-free `hot-spots' (Westphal et al. 1974)
that typically cover about 1\% of the surface area of the planet
(Orton et al. 1996).
Thus the apparent rapid change from cloudy L-dwarfs to clear T-dwarfs may 
be due to a gradual change in cloud coverage in the
visible atmosphere, with the larger flux from the clear regions
quickly dominating.  

\section{Summary}

We have developed a simple cloud model for substellar atmospheres
that includes precipitation by condensate
particles larger than that set by the convective velocity scale,
which permits us to reproduce the properties retrieved
from Jovian ammonia clouds. 
Effective precipitation also produces
cloud profiles in theoretical brown dwarf and extrasolar
giant planet atmospheres that are broadly consistent with observations.

As in the solar system, real clouds in the atmospheres of substellar
objects will likely be neither uniform nor homogeneous, however
we hope that this model will provide a framework for evaluating
the globally-averaged role such clouds play in controlling the
thermal radiative transfer and spectra of brown dwarfs and extrasolar
giant planets.  

\acknowledgments

We thank Sarah Beckmann for detecting anomalous behavior in an early
version of the model.  We also thank Robert West and Kevin Zahnle for providing
helpful comments on the manuscript.
M.S.M. acknowledges support from NASA grants
NAG58919 and NAG59273 as well as NSF grants
AST 9624878 and AST 0086288.

\appendix

\section{Saturation Vapor Pressures} \label{pvap}

For the saturation vapor pressure of ammonia ($e_{\rm s}$, in dyne cm$^{-2}$)
we fit the measurements tabulated in the CRC handbook (Weast, 1971) with 
\begin{eqnarray}
e_{\rm s}({\rm NH_3}) & = & \exp \left( 
    10.53 - \frac{2161}{T} - \frac{86596}{T^2} \right)
\end{eqnarray}
where the temperature is in K.  

For the vapor pressure of water we use the expressions of Buck (1981),
over ice for $T~<~273.16$~K, and over liquid water at greater temperatures:
\begin{mathletters}
\begin{eqnarray}
e_{\rm s}({\rm H_2O,ice})    & = & 6111.5\exp \left(
\frac{ 23.036 T_{\rm c} - T_{\rm c}^2/333.7} { T_{\rm c} + 279.82 } \right) \\
e_{\rm s}({\rm H_2O,liquid}) & = & 6112.1\exp \left(
\frac{ 18.729 T_{\rm c} - T_{\rm c}^2/227.3} { T_{\rm c} + 257.87 } \right)
\end{eqnarray}
\end{mathletters}
where $T_{\rm c}$ is the temperature in degrees Celsius.  These expressions
are unsuitable at $T > 1048$~K, leading to vapor pressures that decrease with
with increasing temperatures.  Hence, at greater temperatures we simply fix
$e_{\rm s} \rm (H_2O) = 6\times10^8 \: \rm dyne \: cm^{-2}$, which is its value
at $T = 1048$~K.

The vapor pressures for iron and enstatite are taken from Barshay and 
Lewis (1976).  For iron below and above its melting point of 1800~K,
we use, respectively
\begin{mathletters}
\begin{eqnarray}
e_{\rm s}({\rm Fe,solid})  & = & \exp \left(15.71 - \frac{47664}{T} \right) \\
e_{\rm s}({\rm Fe,liquid}) & = & \exp \left(9.86 - \frac{37120}{T} \right)
\end{eqnarray}
\end{mathletters}
and for enstatite we use
\begin{eqnarray}
e_{\rm s}({\rm MgSiO_3}) & = & \exp \left( 25.37 - \frac{58663}{T} \right)
\end{eqnarray}

\section{Sedimentation Velocities} \label{vfall}

Droplet terminal fallspeeds are calculated by first
assuming viscous flow around spheres corrected for gas kinetic effects:
\begin{equation}
v_{\rm f} =  \frac {2}{9} \frac { \beta g r^2 \Delta \rho} {\eta}
\end{equation}
where $g$ is the gravitational acceleration, $r$ is the droplet radius,
and $\Delta \rho = \rho_{\rm p} - \rho_{\rm a}$ is the
difference between the densities of the condensate and the atmosphere.
The Cunningham slip factor, $\beta~=~(1~+~1.26 N_{\rm Kn})$, 
accounts for gas kinetic effects,
in which the Knudsen number ($N_{\rm Kn}$)
is the ratio of the molecular mean free path to the droplet radius.
The dynamic viscosity of the atmosphere is given by 
Rosner (2000):
\begin{equation}
\eta =  \frac{5}{16} 
        \frac{ \sqrt{\pi m k_{\rm B} T} }
         { \pi d^2}
         \frac{ \left( k_{\rm B} T / \epsilon \right)^{0.16} }
         { 1.22 }
\end{equation}
where $d$ is the molecular diameter and $\epsilon$ is the depth
of the Lennard-Jones potential well
for the atmosphere ($2.827\times10^{-8}$ cm and $59.7 k_{\rm B}$ K,
respectively, for H$_2$) and $k_{\rm B}$ is the Boltzmann constant.

For turbulent flow, at Reynolds numbers
($N_{\rm Re} = 2 r \rho_{\rm a} v_{\rm f} / \eta $)
between 1 and 1000,
we use a standard trick to solve the drag problem.  Noting that
$C_{\rm d} N_{\rm Re}^2 = 32 \rho_{\rm a} g r^3 \Delta \rho / 3 \eta^2$
is independent of fall velocity,
we fit $y = \log( N_{\rm Re} )$ as a function of
$x = \log( C_{\rm d} N_{\rm Re}^2 )$ to the following data:
at $N_{\rm Re}$ = 1 we assume viscous flow, with $C_{\rm d} = 24$;
for intermediate Reynolds numbers we use the 
data for rigid spheres from Table 10-1 of Pruppacher \& Klett (1978);
and at $N_{\rm Re}$ = 1000 we assume an asymptote of $C_{\rm d} = 0.45$.
This asymptote is appropriate to moderately oblate spheroids
(Figure 10-36 in Pruppacher \& Klett 1978),
which are more appropriate to unknown condensates than the
extreme case of smooth spheres.  Our fit to the data is
$y = 0.8x - 0.01x^2$, which
allows us to evaluate the droplet terminal fall velocity from
$N_{\rm Re}$.

At Reynolds numbers $>$ 1000 we assume the drag coefficient is fixed
at its asymptotic value ($C_{\rm d}~=~0.45$), which leads to
\begin{equation}
v_{\rm f} = \beta \sqrt{ \frac{ 8 g r \Delta \rho } {3 C_{\rm d} \rho_{\rm a}} }
\end{equation}

We assume rigid particles and thereby ignore breakup, for instance
by liquid droplets due to hydrodynamic instability.
For the density of ammonia ice particles, we use 0.84 g~cm$^{-3}$
(Manzhelii \& Tolkachev 1964); 
for water we use 0.93 g~cm$^{-3}$ (corresponding to ice at a temperature of
200 K, using Equation 4-17 from Pruppacher \& Klett 1978); and
for enstatite and iron we use 3.2 and 7.9 g~cm$^{-3}$, respectively
(Table 1.18 of Lodders \& Fegley, 1998).

\clearpage




\begin{deluxetable}{ccc}

\tabletypesize{\scriptsize}
\tablecaption{Adjustable model parameters. \label{table1}}
\tablewidth{0pt}
\tablehead{
\colhead{Parameter} & \colhead{Baseline value}   & \colhead{Description} }

\startdata
$f_{\rm rain}$   & 3 & Ratio of mass-weighted sedimentation velocity to
convective velocity scale \\
$K_{\rm min}$    & $10^5$ cm$^2$ s$^{-1}$ & Minimum value of eddy diffusion
coefficient \\
$\Lambda$        & 0.1 & Minimum ratio of turbulent mixing length to atmospheric
scale height \\
$S{\rm cloud}$   & 0   & Supersaturation that persists after accounting for
condensation \\
$\sigma_{\rm g}$ & 2 & Geometric standard deviation in lognormal size
distributions of condensates
\enddata

\end{deluxetable}

\clearpage

\begin{figure}
\plotone{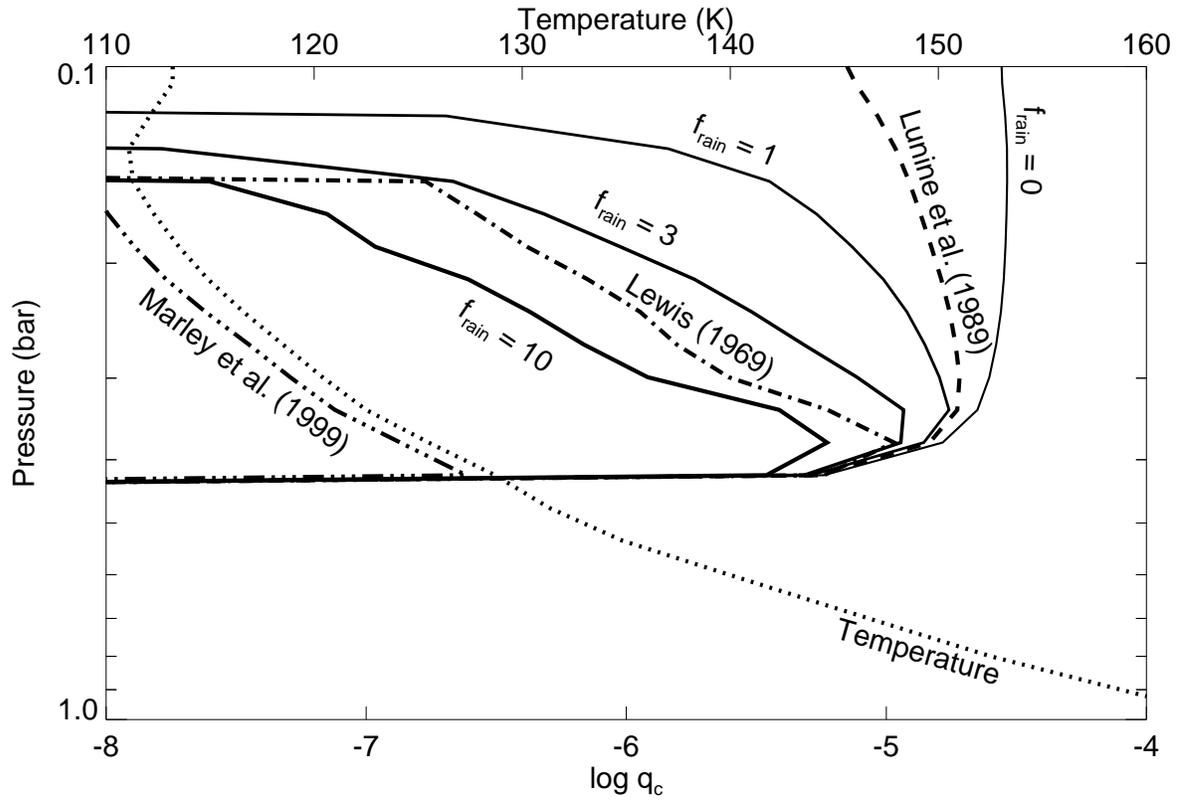}
\figcaption{ \label{fig1}
Vertical profiles of mole fraction (mixing ratio by volume) of condensed
ammonia ($q_{\rm c}$) from present model of Jovian ammonia cloud with
different values $f_{\rm rain}$, and from our adaptations of other models
as labeled.  The vertical coordinate is atmospheric pressure.  The dotted
line is the temperature profile.  The kinks in the condensate profiles are
due to ripples in the temperature profile.  
}
\end{figure}

\clearpage

\begin{figure}
\plotone{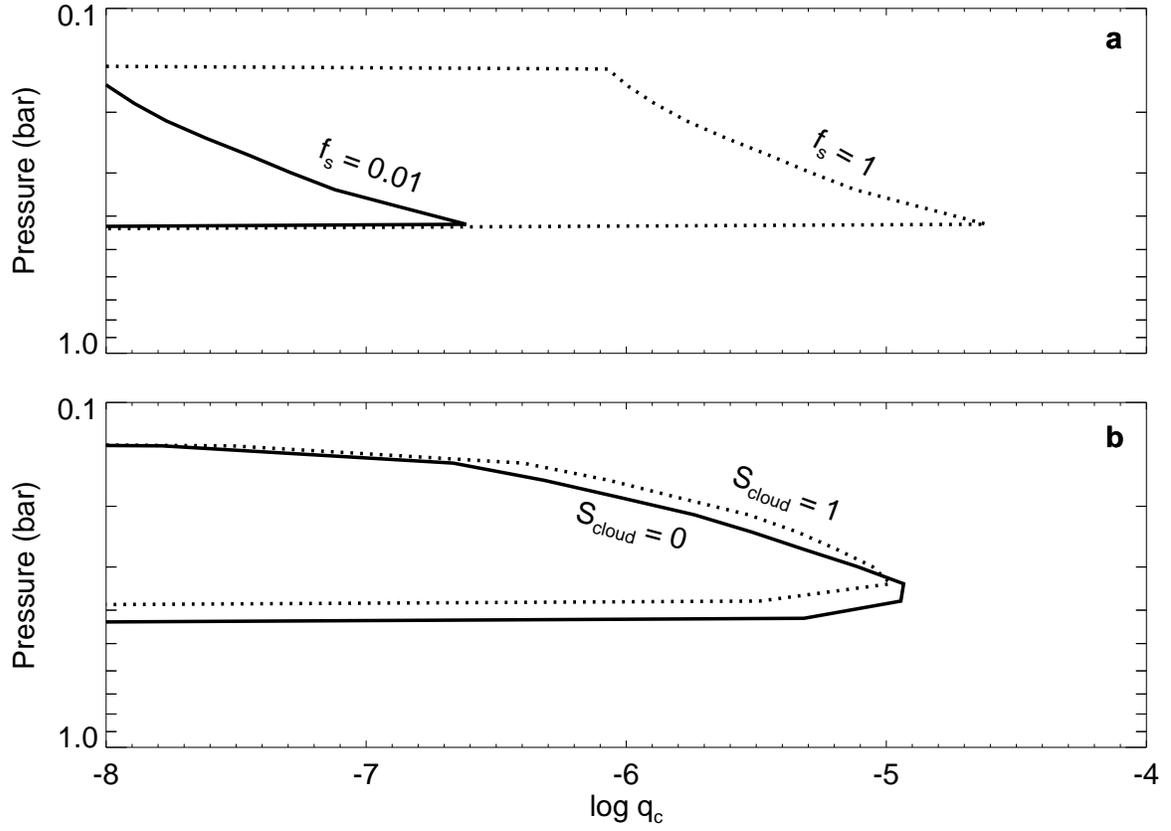}
\figcaption{ \label{fig2}
Vertical profiles of condensed ammonia (as in Figure 1) from
(\emph{a}) the model of Marley et al. (1999) for two values of $f_{\rm s}$
(the potential supersaturation prior to condensation), and
from (\emph{b}) our baseline model for two comparable values of
$S_{\rm cloud}$ (the supersaturation persisting after condensation).
}
\end{figure}

\clearpage

\begin{figure}
\plotone{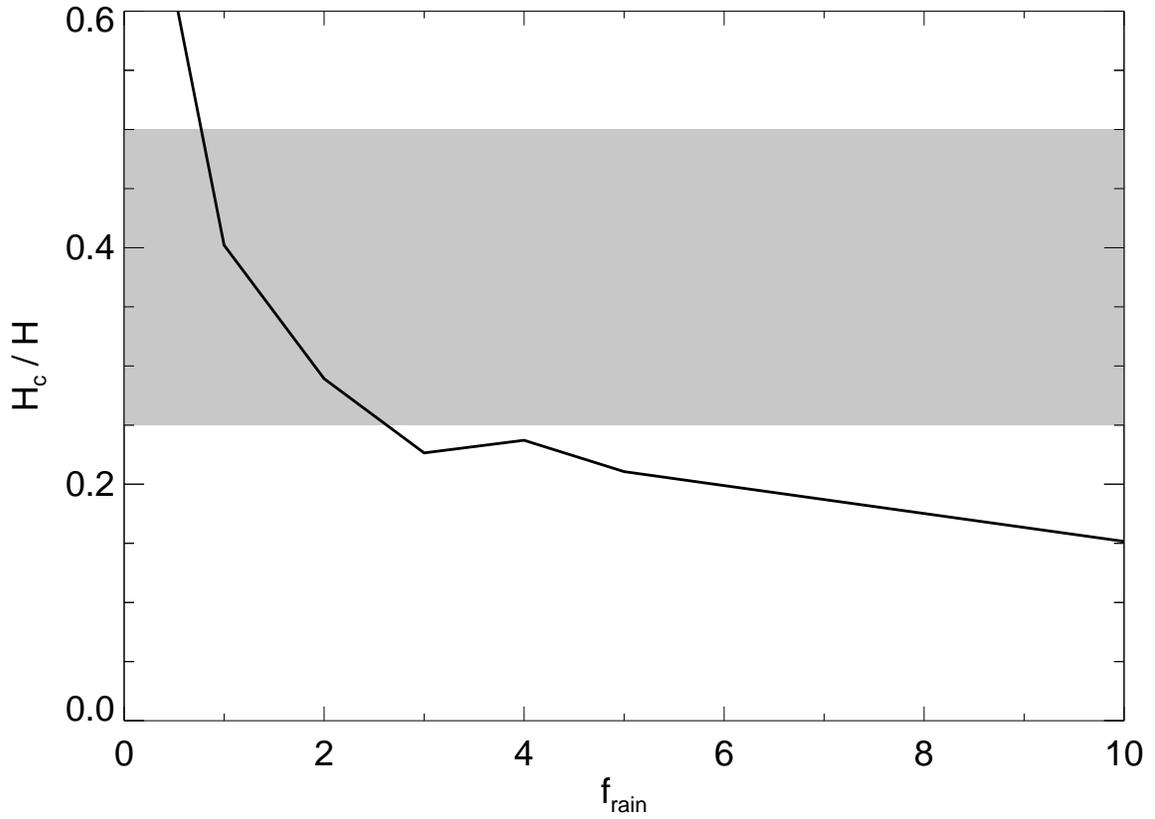}
\figcaption{ \label{fig3}
The ratio of the condensed ammonia scale height 
to the atmospheric pressure scale height ($H_{\rm c}$ and $H$, respectively)
as a function of $f_{\rm rain}$.
The model condensate scale height is calculated from the altitude of the peak
opacity (for geometric scatterers) and the altitude at which the
opacity falls to exp(-1) of its peak.
The grey region depicts the range of
scale height ratios retrieved from the Jovian Equatorial and Northern Tropical
Zones by Carlson et al. (1994).
}
\end{figure}

\clearpage

\begin{figure}
\plotone{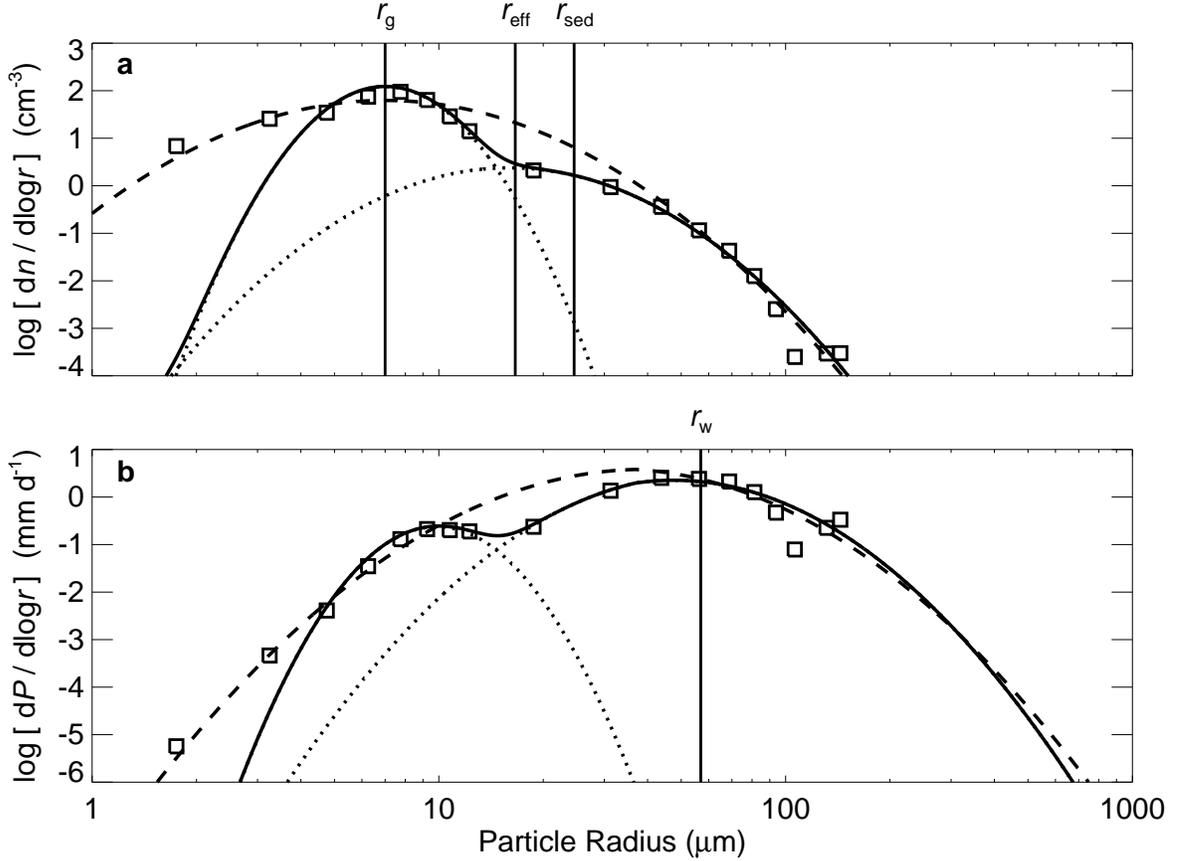}
\figcaption{ \label{fig4}
Measured size distributions (squares) of (\emph{a}) cloud droplet number
concentration ($n$, in cm$^{-3}$) and (\emph{b}) precipitation flux ($P$,
in mm d$^{-1}$) measured in California stratocumulus and averaged over
12 km of flight at 200 m altitude, $\sim$100 m below cloud top
(from Figure 3\emph{b} of Ackerman et al. 2000b). 
Solid curve is a bimodal fit to the measurements, the sum of
two lognormal distributions (dotted curves). Dashed curve is
a monomodal lognormal fit to the measurements, in which
$N$ = 40 cm$^{-3}$, $r_{\rm g}$ = 7 $\mu$m, and $\sigma_{\rm g}$ = 1.8
(symbols defined in text). The values of $r_{\rm g}$, $r_{\rm eff}$, and
$r_{\rm sed}$ are tied to the monomodal fit to the measurements, while
$r_{\rm w}$ is calculated from the convective velocity, measured to be
0.33 m s$^{-1}$ (I. Brooks 1995, private communication).
}
\end{figure}

\clearpage

\begin{figure}
\plotone{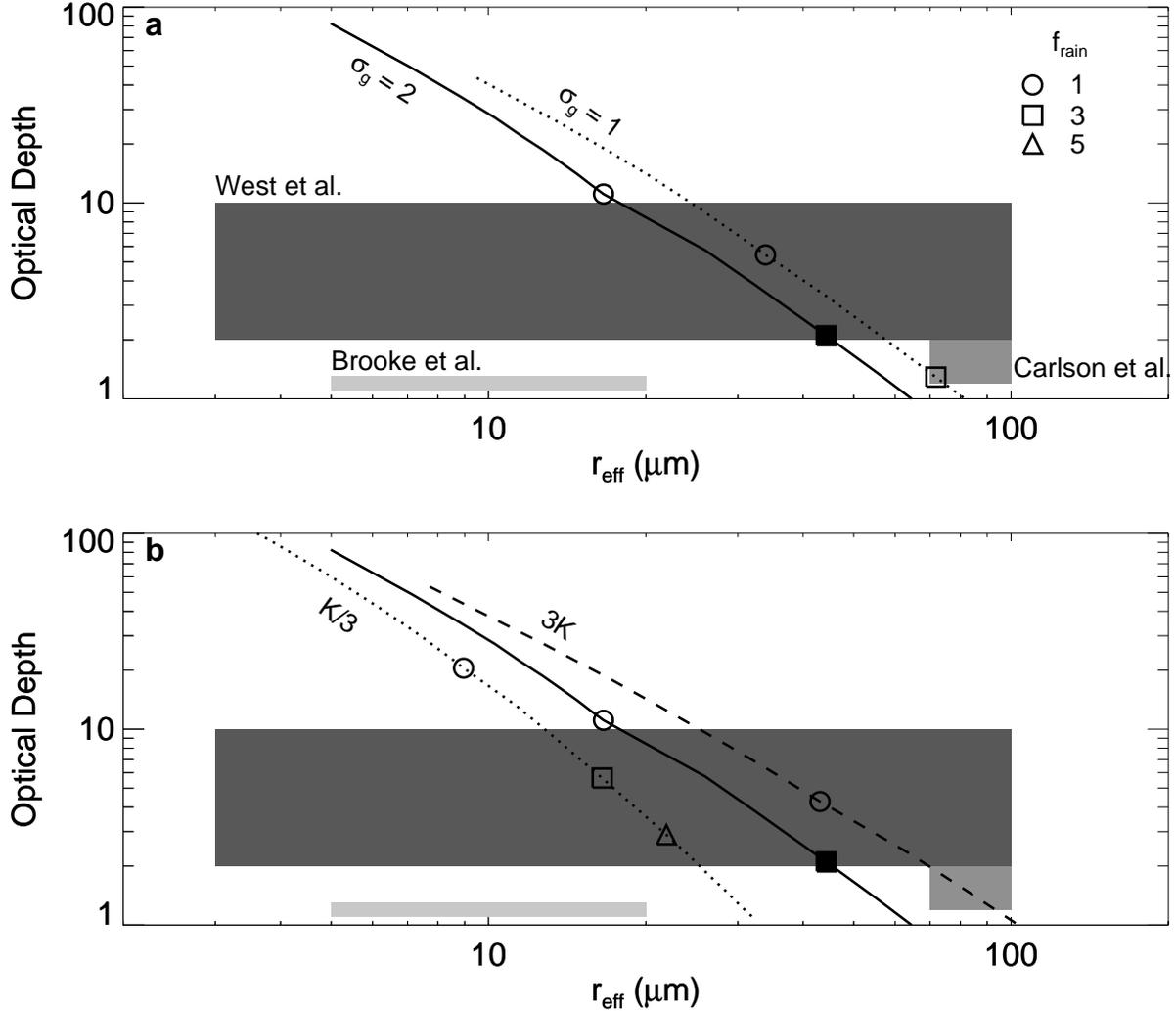}
\figcaption{ \label{fig5}
Condensate optical depth (for geometric scatterers) plotted as a function
of particle effective radius ($r_{\rm eff}$) for the Jovian ammonia
cloud.  The darkest grey region corresponds to the range of observations
given by West et al. (1986) (equivalent to Banfield et al. 1998, as
described in the text), the medium grey region corresponds to the 
retrievals from Voyager IRIS
$5 - 45 \, \mu \rm m$ observations by Carlson et al. (1994), and the
light grey region corresponds to retrievals from ISO 3-$\mu \rm m$
observations by Brooke et al. (1998).
The lines correspond to variation of model results as $f_{\rm rain}$ 
ranges from 0.1 to 10, with results at $f_{\rm rain}$ of 1, 3, and 5
marked by symbols as denoted in (\emph{a}).
The particle effective radii correspond to an opacity-weighted
averages.
The solid lines correspond
to baseline values of (\emph{a}) the width of the log-normal particle
size distribution ($\sigma_{\rm g}$) and (\emph{b}) the eddy diffusion
coefficient ($K$).
The dotted line in (\emph{a}) corresponds to a monodisperse particle
size distribution; the dotted and dashed lines in (\emph{b}) correspond
respectively to dividing and multiplying $K$ by a factor of 3.
Filled-in squares denote model results using the complete baseline set
of parameters.
}
\end{figure}

\clearpage

\begin{figure}
\plotone{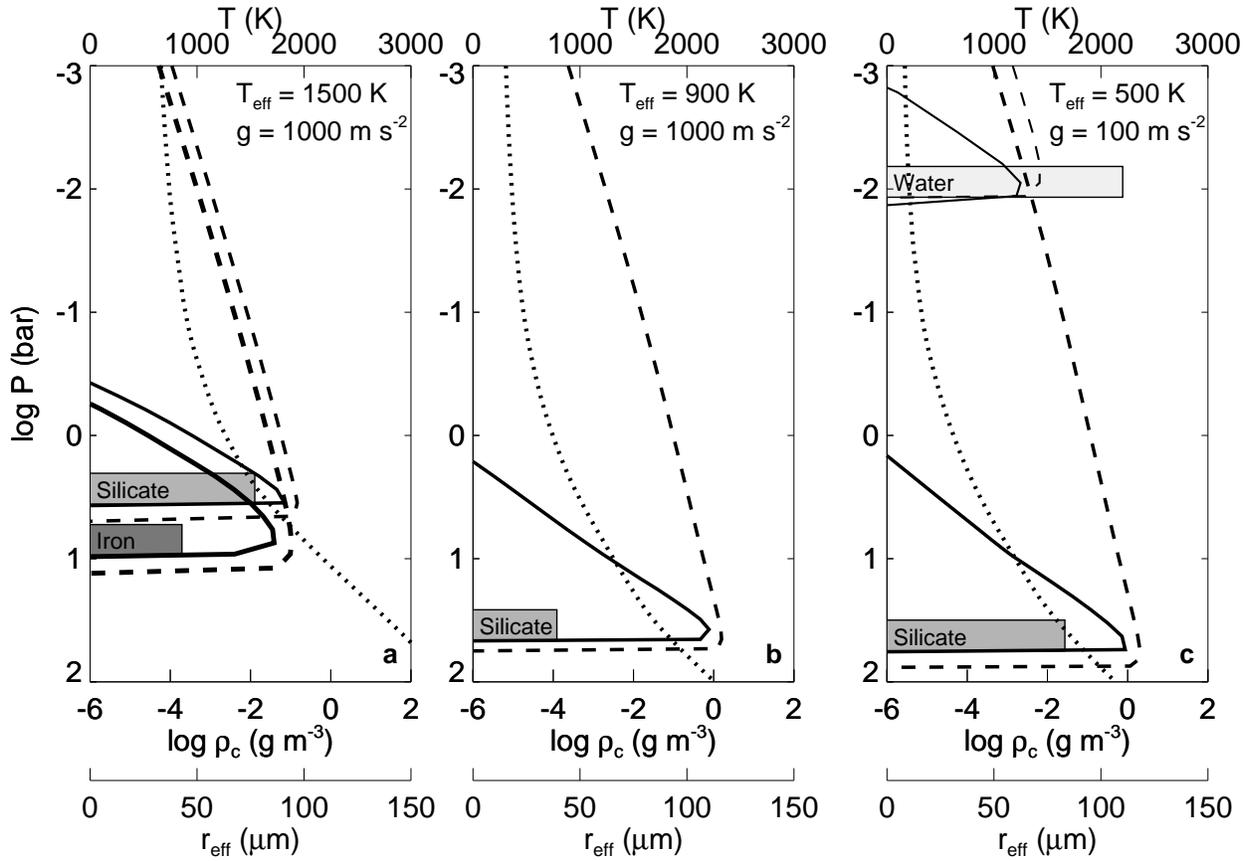}
\figcaption{ \label{fig6}
Profiles of temperature (dotted curves)
and condensate mass concentration from baseline model
($\rho_{\rm c}$, solid curves)
in theoretical atmospheres of a (\emph{a}) L-dwarf, (\emph{b}) T-dwarf,
and (\emph{c}) extrasolar giant planet.  Droplet effective radii at
cloud base are shown as horizontal bars.  Well-mixed clouds are shown
as dashed curves.  The theoretical temperature profiles are calculated
for cloud-free conditions.
}
\end{figure}

\clearpage

\begin{figure}
\plotone{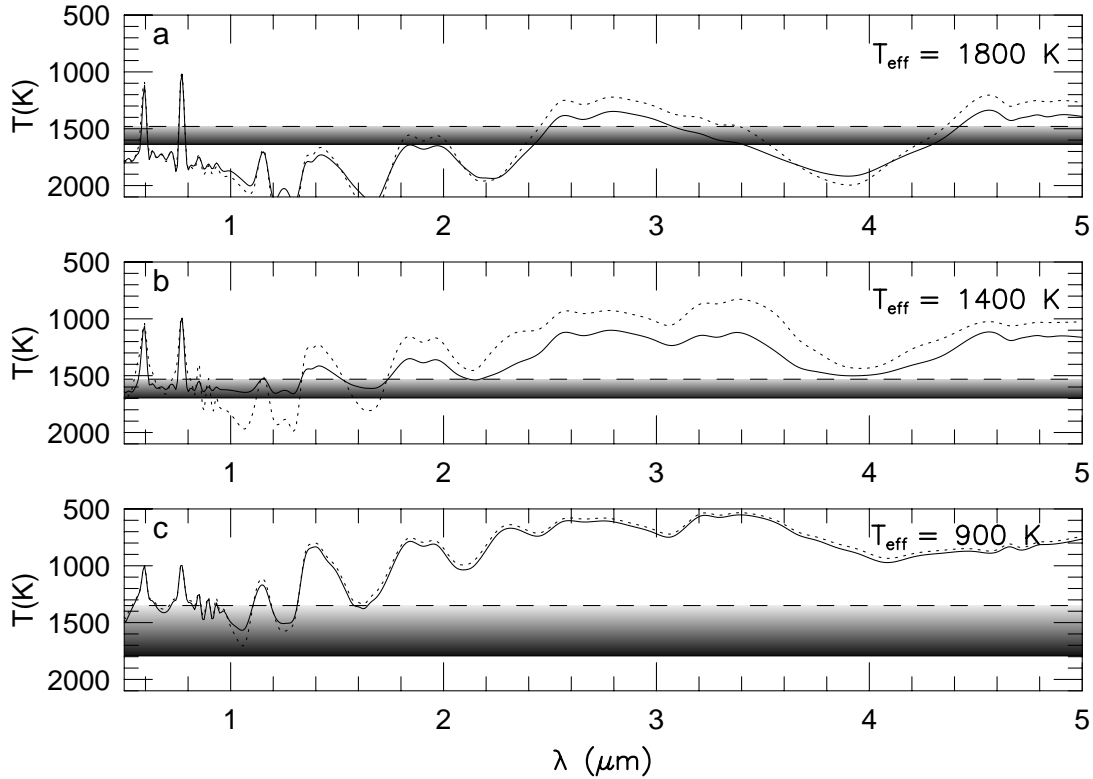}
\figcaption{ \label{fig7}
Brightness temperature as a function of wavelength for atmosphere 
models calculated self-consistently (Marley et al. 2001) to include 
(solid) or exclude (dotted) silicate and iron clouds.  Brightness 
temperature increases downward to indicate increasing depth in the 
atmosphere.  Clouds are calculated using the baseline model 
parameters.  Solid straight line indicates base of 
silicate cloud, dashed line denotes level in atmosphere at which 
column extinction optical depth reaches 0.1, and shading depicts
the decrease in cloud extinction with altitude.  Since cloud 
particle radius exceeds 10 $\mu$m in these models, the Mie 
extinction efficiency is not a strong function of wavelength over 
the range shown.  Shown are models characteristic of (\emph{a}) an early-type 
L-dwarf with $T_{\rm eff} = 1800\,\rm K$, (\emph{b}) a late L with
$T_{\rm eff} = 1400\,\rm K$, and (\emph{c}) a T-dwarf with
$T_{\rm eff} = 900\,\rm K$.  All atmosphere models are for solar
composition and gravitational acceleration of 1000 m s$^{-2}$,
roughly appropriate for a 30 Jupiter-mass brown dwarf.  Sodium and potassium 
lines, calculated using the theory of Burrows, Marley, and Sharp 
(2000), are prominent in the optical.
}
\end{figure}

\end{document}